\documentclass[preprint,aip,graphicx]{revtex4-1}
\usepackage{amsmath,amsbsy}
\usepackage{amsfonts}
\usepackage{amssymb}
\usepackage{lineno}
\usepackage{graphicx}
\usepackage{mathrsfs}
\usepackage[final]{changes}

\def\e{\mathrm{e}}
\def\d{\mathrm{d}}
\def\ii{\mathrm{i}}
\def\bx{\boldsymbol{x}}
\def\bn{\boldsymbol{n}}
\def\bX{\boldsymbol{X}}
\def\br{\boldsymbol{r}}
\def\bv{\boldsymbol{v}}
\def\eps{\epsilon}
\def\be{\boldsymbol{e}}
\def\bF{\boldsymbol{F}}

\newcommand{\av}[1]{\left\langle #1 \right\rangle }
\newcommand{\rb}[1]{\left( #1 \right) }
\newcommand{\bs}[1]{{\boldsymbol{#1}}}

\newcommand*\patchAmsMathEnvironmentForLineno[1]{%
  \expandafter\let\csname old#1\expandafter\endcsname\csname #1\endcsname
  \expandafter\let\csname oldend#1\expandafter\endcsname\csname end#1\endcsname
  \renewenvironment{#1}%
     {\linenomath\csname old#1\endcsname}%
     {\csname oldend#1\endcsname\endlinenomath}}%
\newcommand*\patchBothAmsMathEnvironmentsForLineno[1]{%
  \patchAmsMathEnvironmentForLineno{#1}%
  \patchAmsMathEnvironmentForLineno{#1*}}%
\AtBeginDocument{%
\patchBothAmsMathEnvironmentsForLineno{equation}%
\patchBothAmsMathEnvironmentsForLineno{align}%
\patchBothAmsMathEnvironmentsForLineno{flalign}%
\patchBothAmsMathEnvironmentsForLineno{alignat}%
\patchBothAmsMathEnvironmentsForLineno{gather}%
\patchBothAmsMathEnvironmentsForLineno{multline}%
}

\begin{document}


\title{Dynamics of a spherical particle in an acoustic field: a multiscale approach}
\author{Jin-Han Xie}
\email{J.H.Xie@ed.ac.uk}
\author{Jacques Vanneste}
\affiliation{ School of Mathematics and Maxwell Institute for Mathematical Sciences,
University of Edinburgh, Edinburgh EH9 3JZ, UK}
\date{\today}




\begin{abstract}
A rigid spherical particle in an  acoustic wave field oscillates at the wave period  but  has also a mean motion on a longer time scale. The dynamics of this mean motion is crucial for numerous applications of acoustic microfluidics, including particle manipulation and flow visualisation.  It is controlled by four physical effects: acoustic (radiation) pressure, streaming, inertia and viscous drag. In this paper, we carry out a systematic multiscale analysis of the problem in order to assess the relative importance of these effects depending on the parameters of the system that include wave amplitude, wavelength, sound speed, sphere radius, and viscosity. 

We identify two distinguished regimes characterised by a balance among three of the four effects, and we derive the equations that govern the mean particle motion in each regime. This recovers and organises classical results by King, Gor'kov and Doinikov, clarifies the range of validity of these results, and reveals a new nonlinear dynamical regime.  In this regime, the mean motion of the particle remains intimately coupled to that of the surrounding fluid, and while viscosity affects the fluid motion, it plays no part in the acoustic pressure. 
Simplified equations, valid when only two physical effects control the particle motion, are also derived. They are used to obtain sufficient conditions for the particle to behave as a passive tracer of the Lagrangian-mean fluid motion.

\end{abstract}

\maketitle 

\section{Introduction}

High-frequency acoustic waves are increasingly used to actuate and manipulate fluids at microscales, with applications that include flow generation, \citep{From2008,Algh2011} particle collection or separation,\citep{Shir1973,Hu2004,Li2007,Ober2009,Roge2010,Barm1982,Gros1998} acoustic levitation, \citep{vand2005} and calibration of high-frequency transducers \citep{Chen1996}. As a result, the field of acoustic microfluidics is in rapid development (see Refs.\ \citenum{Frie2011,Yeo2013} for recent reviews). Many applications and experiments involve the motion of small, typically spherical particles, either as objects to be manipulated, or simply as tracers used to visualise the flow. There is therefore considerable interest in modelling the dynamics of such particles, especially the mean dynamics that results from averaging over many wave periods. This is a classical problem that has motivated a great deal of work which we review briefly below. (see also the review in Ref \citenum{Bruu2012}) Most of this work focuses on the calculation of the acoustic (radiation) pressure acting on particles as a result of wave scattering. This is not the only physical effect affecting the motion of particles. Acoustic streaming\citep{Nybo1965,Ligh1978,Rile2001} -- the nonlinear generation of a mean flow by acoustic waves -- acts even in the absence of particles and is the main mechanism exploited in acoustic microfluidics; it clearly affects the dynamics of particles, though in a way that is sometimes difficult to distinguish from acoustic radiation. \citep{Dani2000} The inertia of the particles and of the fluid, and the viscous (Stokes) drag caused by the motion relative to the surrounding fluid are the other two main physical effects.  

The relative role of these four effects -- acoustic pressure, streaming, inertia and viscous drag -- is the main theme of this paper. Specifically, we explore how, depending on the parameters of the problems, the mean dynamics of a single spherical rigid particle can be controlled by different balances among these effects, and we derive the corresponding mean equations of motion. We do so by applying a systematic multiscale approach: taking the standard linear acoustics hypothesis of small-amplitude waves, characterised by an acoustic Mach number $\eps \ll 1$, we consider possible distinguished scalings of the other parameters in the problem, primarily viscosity measured by the ratio $\delta/a$ of the Stokes boundary-layer thickness to the sphere radius. We apply multiscale and matched-asymptotics methods to obtain asymptotic equations governing the motion over long time scales. Crucially, this requires (i) to consider the fully coupled fluid-particle system, recognising that a reduction to an ordinary differential equation for the particle alone is possible only in certain parameter regimes; and (ii) to take into account explicitly the mean displacements of the particle. This is in contrast with much of previous work which, as mentioned, concentrates on acoustic pressure and typically assumes that the mean particle position can be taken as frozen. 

In the classic work by \citet{King1934} and \citet{Gork1962} on acoustically-driven particles in inviscid fluids (with assumptions of, respectively, axisymmetry and long wavelength $ka \ll 1$, with $k$ the wavelength), explicit expressions are derived for the acoustic pressure in this way; the effect of the fluid motion is subsequently taken into account in a somewhat ad hoc manner by including an added mass effect in Newton's second law for the particle. Our treatment shows this to be valid for sufficiently small viscosity and makes precise how small the viscosity needs to be. The case of a viscous fluid has been considered in many papers,\citep{west1951,Dani1985,Dani1986}
culminating in the work of Doinikov \citep{Doin1994A,Doin1994J,Doin1994R} who provides complete expressions for the acoustic pressure in an axisymmetric field for arbitrary viscosity and wavelength. Simplified expressions valid for $ka \ll 1$ and arbitrary wave fields have recently been obtained by \citet{Settnes12} (see also Ref.\ \citenum{Dani2000} and references therein). In this viscous case, a closed equation of motion for the sphere can be inferred from the acoustic force by assuming that the inertia of the particle and surrounding fluid is negligible. Again, our treatment shows this to be a valid approximation under conditions that we make explicit. More importantly, our analysis reveals a new regime (termed Regime II below) in which the fluid motion driven by the particle is both crucial for the particle dynamics and determined by the full (viscous) Navier--Stokes equations instead of the simple potential solution relevant in the purely inviscid approximation. In this regime,  the particle and fluid motion are completely coupled, and no reduction to a single ordinary differential equation is possible.

Much of the earlier work on acoustic pressure was motivated by applications very different from those arising today from developments in acoustic microfluidics. The focus of this paper reflects these developments: in particular, we pay attention to the case of particles with the same density as the fluid. While this case is `of no interest for practice' \citep{Doin1994R} when dealing with, say, dust particles or water drops in air, it is highly relevant in microfluidics applications where the particle density is often selected to avoid buoyancy effects. We examine the conditions that need to be satisfied for such particles to follow fluid elements and hence act as genuinely passive tracers. This is important in view of the widespread use of particles for this purpose in acoustic microfluidics.

The paper is structured as follows. Section \ref{strategy} introduces the governing equations and relevant non-dimensional parameters. Based on this, and under the assumption $ka = O(1)$ (which includes $ka \ll 1$), it gives a heuristic argument for the existence of two distinguished asymptotic regimes in which three of the four physical effects affecting particle motion balance. These two regimes are considered in detail in sections \ref{RegI} and \ref{RegII}. There we apply systematically multiscale asymptotics to derive the equations governing the mean dynamics in each regime. These equations can be further simplified in several intermediate regimes in which only two of the four physical effects come into play. These regimes are of great practical importance; the relevant equations are derived in section \ref{IntReg}. The paper concludes in section \ref{Dis} with a brief summary, a discussion of the relevance of the results to examples of acoustic microfluidics experiments, and pointers to further work, including on the case $ka \gg 1$. Throughout, we emphasise the systematic derivation of  mean equations of motion over  specific expressions for the acoustic-pressure terms. We refer the reader to earlier work for these and point out the approximations that can be made consistently in each of the regimes we analyse. We note that complete expressions for the acoustic pressure have been obtained for axisymmetric wave fields (Ref.\ \citenum{Doin1994R} and references therein) and for general wave fields provided that $ka \ll 1$ (Refs.\ \citenum{Dani2000,Settnes12} and references therein). Our results have the same range of validity.

\section{Formulation}\label{strategy}

\subsection{Dimensionless parameters and scaling}

\begin{figure}
\centering
\includegraphics[width=8cm]{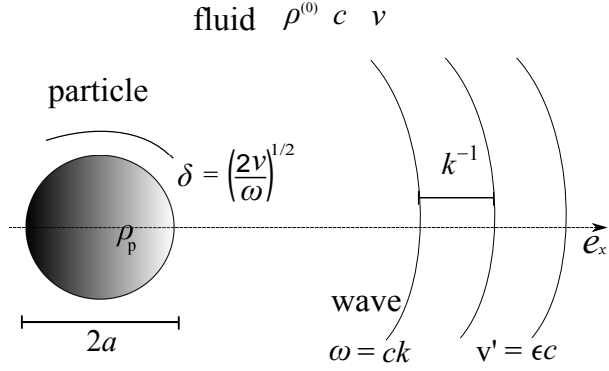}
\caption{Parameters controlling the motion of a spherical particle in an acoustic field.}
\label{system}
\end{figure}

We study the mean dynamics of a rigid sphere in an axisymmetric acoustic wave field. For simplicity we neglect the effect of heat conduction \citep{Doin1997I,Doin1997II}. The problem is then characterised by 8 parameters: the fluid properties determine the equilibrium density $\rho^{(0)}$, sound speed $c$ and shear and bulk viscosities $\eta$ and $\xi$; the particle is characterised by its density $\rho_\mathrm{p}$ and radius $a$;
the incident wave by  a frequency $\omega$ and a velocity amplitude $v'$. See Figure \ref{system} for an illustration.
The $\pi$-theorem of dimensional analysis  yields 5 dimensionless parameters: $\eta/\xi$,
$\lambda =\rho^{(0)}/\rho_\mathrm{p}$, 
$\epsilon = v'/c$, 
$\delta/a$, where $\delta = \sqrt{2 \eta /(\rho^{(0)} \omega)} = \sqrt{2\nu/\omega}$ is the Stokes boundary-layer thickness (e.g., Ref.\ \citenum{Batc1967}, section 5.13), and $ka$, where $k=\omega/c$ is wavenumber.  

Since we are dealing with acoustic waves, we naturally assume that the acoustic Mach number $\eps$ is small: $\epsilon \ll 1$. Different dynamical regimes then emerge depending on the size of the other dimensionless parameters relative to $\eps$. We assume that both $\eta/\xi$ and $\lambda$ are $O(1)$ as is relevant to most applications. This leaves the two parameters $\delta/a$ and $ka$ which we relate to $\eps$ according to
\begin{equation} \label{alphagamma}
\delta/a = O(\epsilon^\alpha) \quad \textrm{and} \quad  k a = O(\epsilon^\gamma).
\end{equation}
The exponents $\alpha$ and $\gamma$ introduced in (\ref{alphagamma}) control the nature of the dynamics. Their physical interpretation is clear: increasing $\alpha$ decreases the strength of the viscous effects, while increasing $\gamma$ increases the wavelength. 

The problem at hand involves two distinct time scales: the wave time scale $\omega^{-1}$ and a slower time scale characterising the mean motion of the sphere. To capture this, we introduce a slow time variable $T$, related to the fast (wave) time $t$ by
\begin{equation} \label{T}
T=\epsilon^{\beta} t.
\end{equation}
To apply systematic asymptotic methods, the exponent $\beta$ should be related to $\alpha$ and $\gamma$. This requires to consider to the balance of terms in the equations governing the dynamics of the coupled fluid-particle system.

\subsection{Basic equations}

The fluid is governed by the compressible Navier--Stokes equations
\begin{subequations} \label{navier}
\begin{align}
\frac{\partial }{\partial t}(\rho \bv) &= \nabla \cdot  (\sigma - \rho \bv\otimes\bv ),\\
\frac{\partial \rho}{\partial t} + \nabla \cdot (\rho \bv) &= 0, \label{012} \\
\textrm{with} \quad \sigma_{ij} &= -p\delta_{ij}+\eta \left( \frac{\partial v_i}{\partial x_j} + \frac{\partial v_j}{\partial x_i} - \frac{2}{3} \frac{\partial v_k}{\partial x_k}\delta_{ij} \right) + \xi \frac{\partial v_k}{\partial x_k}\delta_{ij}, \label{015}
\end{align}
\end{subequations}
where $\rho$ is the fluid density, $\bv$ its velocity, $\sigma$ is the stress tensor, and $p$ is the pressure. Subscripts denote components and Einstein's summation convention is used. \replaced{Here we neglect thermodynamic effects and assume that the fluid is barotropic (e.g., homentropic) so that Eqs.\ (\ref{navier})  
are supplemented by an equation of state $p=p(\rho)$. The boundary conditions are given by the no-slip boundary condition at the surface of the particle}{Eqs.\ (\ref{navier}) are supplemented by an equation of state $p=p(\rho)$, by the no-slip boundary condition at the surface of the particle}
\begin{equation}
\bv(\mathbf{\bx},t) = \frac{\d \bX}{\d t} \quad  \textrm{for} \ \ \mathrm{\bx} \in S_{\bX}, \label{bc}
\end{equation}
where $\bx$ is the position vector, $\bX$ the position of the centre of the particle, and $S_{\bX}$ denotes the sphere of radius $a$ centred at $\bX$, and by a prescribed incident acoustic field at infinity 
\begin{equation}
\bv( \bx,t ) \sim \bv_\mathrm{incident} \quad \textrm{as} \ \ \mathbf{\bx} \to \infty.
\end{equation}
The motion of the particle is governed by Newton's second law written as
\begin{equation}
 M \frac{\d^2 \bX}{\d t^2} = \int_{S_{\bX}} \sigma \cdot  \bn \, \d s , \label{newton}
\end{equation}
where $M=4\pi \rho_{\mathrm{p}} a^3/3$ is the mass of the particle and $\bs{n}$ denotes the outer normal.
Since the sphere is symmetric and placed in an axisymmetric wave field, its motion is one dimensional along the axis of symmetry of the wave field; we choose this direction to be the $x$ axis, with unit vector $\be_x$, so that $\bX = X \be_x$. \added{We do not take the spin of the sphere into account since none is induced by an axisymmetric wave field.}
Note that the assumption of an axisymmetric wave can be relaxed when $ka \ll 1$ since an arbitrary  wave field is can then be regarded as locally planar. In view of the practical importance of this approximation, satisfied in the majority of applications, we write our results, whenever possible, in a vector form that can be employed for general wave fields when $ka \ll 1$.

\subsection{Regimes} \label{regimes}

Our focus is on the mean motion of the particle, driven by the force on the right-hand side of (\ref{newton}) averaged over a wave period. We denote this average by $\av{\cdot}$, so that $\av{\partial_t \cdot}=0$. 
Considering an expansion of all the variables in the form
\begin{subequations} \label{expand}
\begin{align}
\bv&=\epsilon \bv^{(1)} +\epsilon^2 \bv^{(2)} + \cdots, \label{vexpand} \\
\bx &= \bX^{(0)} + \eps \bX^{(1)} + \eps^2 \bX^{(2)} + \cdots,  \label{xexpand}
\end{align}
\end{subequations} 
where $\bX^{(0)}$ captures the mean motion of the particle,
we write the averaged force as
\begin{equation}
{\left\langle \bs{F} \right\rangle } =  \int_{S_{\bX^{(0)}}} \left\langle \sigma^{(2)}\right\rangle  \cdot \bs{n} \,  \d s - \int_{S_{\bX^{(0)}}} \rho^{(0)} \left\langle \bv^{(1)}\otimes \bv^{(1)}\right\rangle  \cdot \bs{n} \, \d s  + O(\epsilon^3). \label{totforce}
\end{equation}
Note that the integrations are over the surface of the sphere centred at $\bX^{(0)}$ which moves only over the slow time scale. The second term on the right-hand side, however, arises from the integration of $\sigma^{(1)}$ over the rapidly moving surface $S_{\bX^{(0)} + \eps \bX^{(1)}}$ (see, e.g., Ref.\ \citenum{Doin1994R} for a derivation).

The force in (\ref{totforce}) contains two distinct physical effects. The first is viscous drag which relaxes the particle's velocity to the velocity of the surrounding fluid.  It is not obvious what the relevant fluid velocity is but it certainly includes the streaming velocity that is generated by dissipation and nonlinearity even in the absence of a particle \citep{west1951,Nybo1953,Rile2001}. The second effect is the radiation pressure associated with the scattered wave. With this in mind, we can postulate a heuristic form for the equation governing the mean motion of the particle, estimate the order of magnitudes of its terms, and find the combinations of $\alpha$, $\beta$ and $\gamma$ that lead to distinguished limits as $\eps \to 0$.  
These limits are crucial since the corresponding regimes include all the physical mechanisms that can possibly have a leading-order effect simultaneously. Our aim is to derive mean equations that apply to these regimes; simpler models, valid in intermediate regimes, can then be deduced straightforwardly by neglecting certain terms.

Heuristically, we can expect the mean motion of the particle to be governed by an equation of the form
\begin{equation}
\begin{aligned}
				& \tilde{M} \ddot \bX^{(0)} & + & 6\pi a\eta (\dot \bX^{(0)} & -& ~~\tilde{\bv} )&=~~& \bF_\mathrm{ap},\\
\textrm{relative~ order:}~~  & ~~~~\epsilon^{2\beta} & & ~~~~~~~~\epsilon^{2\alpha+\beta} & & \epsilon^{2\alpha+2} & & \epsilon^{2} \label{guessdyeq}
\end{aligned}
\end{equation}
where $\tilde{M}$ is a mass, expected to be the mass of the particle plus a possible added mass stemming from fluid motion,  $\tilde{\bv}$ is a streaming velocity, and $\bF_\mathrm{ap}$ is the acoustic-pressure force. Here and henceforth, the overdot denotes time derivative with respect to the slow time $T$.
Below each term in (\ref{guessdyeq}) we indicate its relative order of magnitude, based on the dimensional estimates $\rho_\mathrm{p} a^3k^{-1}T^{-2}$, $a \eta k^{-1}T^{-1}$, $a \eta v'^2c^{-1}$ and $\rho^{(0)} k a^3v'^2$. These assume: typical particle displacements $\bX^{(0)}$ of size $O(k^{-1})$; a streaming velocity $\tilde \bv = O(v'^2/c)$, which holds provided that the amplitude of the waves varies on an `outer scale' that is not too dissimilar to $k^{-1}$ (e.g., Refs.\ \citenum{Nybo1953,Vann2011}); and the scaling  $k a^3  v'^2$ for the surface integrals in (\ref{totforce}). The latter scaling follows from applying the divergence theorem and the assumption that $ka = O(1)$, i.e., $\gamma \ge 0$ so that the spatial derivatives scale like $k$ rather than $a^{-1}$; it is confirmed by explicit computations of the acoustic radiation force, e.g. in Ref.\ \citenum{Gork1962}. For now, we focus on the case $\gamma \ge 0$ and leave a brief analysis of the case $\gamma < 0$ for Section \ref{sphfol}.


Distinguished limits are obtained by selecting $\alpha$ and $\beta$ to balance as many of the 4 terms in (\ref{guessdyeq}) as possible. It is easy to see that 3 terms at most can be involved in any dominant balance. This yields four possibilities:
\begin{enumerate}
\item The particle's inertia is negligible, leading to the condition $2\alpha+\beta=2\alpha+2=2$, that is, $\alpha=0$ and $\beta=2$, for the balance of the remaining terms. We refer to the corresponding regime  as Regime I.
\item The streaming velocity $\tilde{\bv}$ is negligible, leading to
 $2\beta=2\alpha+\beta=2$, that is, $\alpha=1/2$ and $\beta=1$.
We refer to this as Regime II.
\item The particle's viscous drag is negligible.
Balancing the remaining terms leads to $\alpha=0$ and $\beta=1$ but also to an $O(\eps)$ viscous drag, thus much larger than the other terms, inconsistent with our assumption. There is, therefore, no distinguished limit in which the viscous drag is negligible.
\item Acoustic pressure is negligible. This leads to $\alpha=1$ and $\beta=2$ and again to an inconsistency: there is no distinguished limit with negligible acoustic pressure.
\end{enumerate}
We derive the average equations holding in Regimes I and II in sections \ref{RegI} and \ref{RegII}. Intermediate regimes, in which only two of the terms in (\ref{guessdyeq}) enter the dominant balance, are examined in section \ref{IntReg}. We emphasise that these intermediate regimes, though they may formally correspond to values of $\alpha$ and $\beta$ different from those in Regimes I and II, 
can be deduced as limiting cases. For instance, a balance between the particle's inertia and the acoustic pressure, obtained for $\beta=1$ and any value $\alpha>1/2$, is deduced from Regime II by neglecting viscous drag.

\section{Regime I}\label{RegI}

We first consider the distinguished limit in which the particle's inertia is negligible, corresponding to $\alpha=0$ and $\beta=2$. Because viscosity is an $O(1)$ effect in this case, the averaged force (\ref{totforce}) is given by the complete expression computed by \citet{Doin1994R}. Although he assumed that the particle does not have a slow motion, including this motion requires only a straightforward modification of his calculation because of the large time-scale separation implied by $\beta=2$; the present section is therefore largely a brief review of Ref.\ \citenum{Doin1994R} to which the reader is referred for details. 

\subsection{Wave dynamics}

All dynamical variables are expanded in powers of $\eps$ according to (\ref{expand}) and regarded as functions of both times $t$ and $T$, except for the constant $\rho^{(0)}$ and the $t$-independent $\bX^{(0)}$. We  assume that   $\bX^{(0)}$ captures the entirety of the slow motion so that $\langle \bX^{(j)} \rangle = 0$ for $j \ge 1$. Introducing the expansion into the governing equations 
(\ref{navier})--(\ref{newton}) yields a linear viscous wave equations for $\bv^{(1)}$, $p^{(1)}$ and $\rho^{(1)}$ coupled with the equations 
\begin{equation}
\bv^{(1)} = \partial_t\bX^{(1)} \ \   \textrm{for} \ \ \mathrm{\bx} \in S_{\bX^{(0)}} \quad \textrm{and} \quad 
M \partial_t \bX^{(1)} = \int_{S_{\bX^{(0)}}} \sigma^{(1)} \cdot \bs{n} \, \d s \label{coupling}
\end{equation}
governing the particle motion and its interaction with the fluid. The solution for an axisymmetric flow is best written using spherical polar coordinates centred at $\bX^{(0)}$. With $\theta$ denoting the angle about the axis $\bs{e}_x$, the potential of the incident part of the wave field can be written as
\begin{equation}
\phi_\mathrm{i} = \e^{-\ii \omega t} \sum_{n=0}^\infty A_n(\bX^{(0)}) j_n(kr) P_n(\cos \theta), \label{incexp0}
\end{equation} 
where $r = |\bx - \bX^{(0)}|$, $j_n$ denotes the spherical Bessel function of order $n$, $P_n$ denotes the Legendre polynomial of degree $n$, 
\begin{equation} \label{k} 
k = \omega/(c^2 - \ii \omega (\xi + 4\eta/3)/\rho^{(0)})^{1/2}
\end{equation}
is the wavenumber, and the real part is implied. The amplitudes $A_n$ are determined by the prescription of the incident wave as $r \to \infty$. The scattered part of the wave field is determined by a potential $\phi_\mathrm{s}$ and streamfunction $\psi_\mathrm{s}$ such that $\bv^{(1)}_\mathrm{s} = \nabla \phi_\mathrm{s} + \nabla \times ( \bs{e}_\varphi \psi_\mathrm{s})$, with $\bs{e}_\varphi$ the azimuthal unit vector of the spherical coordinate system; these are given by
\begin{align}
\phi_\mathrm{s} &=  \e^{-\ii \omega t} \sum_{n=0}^\infty  \alpha_n A_n(\bX^{(0)}) h_n(kr) P_n(\cos \theta) \\ \textrm{and} \quad \psi_\mathrm{s} &= \e^{-\ii \omega t} \sum_{n=0}^\infty  \beta_n A_n(\bX^{(0)}) h_n((1+\ii)r/\delta) P_n^1(\cos \theta), 
\end{align}
where $h_n$ is the spherical Hankel function and $P_n^1$ the associated Legendre polynomial. \citet{Doin1994R} gives explicit expressions for the constants $\alpha_n$ and $\beta_n$.

\subsection{Mean dynamics} \label{mean I}

Continuing the expansion of (\ref{navier}) to $O(\eps^2)$ leads to the equations
\begin{subequations} \label{001}
\begin{align}
\rho^{(0)} \partial_t \bv^{(2)} + \partial_t \rb{\rho^{(1)}\bv^{(1)}} &= \nabla \cdot \rb{\sigma^{(2)} - \rho^{(0)} \bv^{(1)} \otimes \bv^{(1)}}, \\
\partial_t \rho^{(2)} &= - \nabla \cdot  \rb{\rho^{(0)} \bv^{(2)} + \rho^{(1)}\bv^{(1)}},  
\end{align}
\end{subequations}
which become
\begin{subequations}  \label{017}
\begin{align}
\nabla \cdot  \av{\sigma^{(2)}}  &= \rho^{(0)} \nabla \cdot \av{\bv^{(1)} \otimes \bv^{(1)}},  \\
\rho^{(0)} \nabla \cdot \av{\bv^{(2)}}  &= -\nabla \cdot \av{\rho^{(1)}\bv^{(1)}}, 
\end{align}
\end{subequations}
upon averaging. Averaging the boundary condition (\ref{bc}) gives
\begin{equation}
\av{\bv^{(2)}} +  \av{\bs{X}^{(1)} \cdot  \nabla \bv^{(1)}} = \dot{\bX}^{(0)} \ \ \textrm{for} \ \  \bx \in S_\mathrm{\bX^{(0)}} \label{mbcI}.
\end{equation}
The left-hand side of (\ref{mbcI}) can be recognised as a Lagrangian mean velocity summing  Eulerian mean velocity and Stokes drift \citep[e.g.][]{Buhl2009}: indeed, in view of (\ref{coupling}), $\bs{X}^{(1)}$ is both the particle and  fluid displacement.
The final equation for the mean flow is provided by the average of (\ref{newton}) to order $O(\eps^2)$. With $\beta=2$, the slow acceleration of the particle is $\eps^4 \ddot \bX^{(0)}$ and does not appear at this order, leaving the dominant balance
\begin{equation}
 \int_{S_{\bX^{(0)}}} \left\langle \sigma^{(2)}\right\rangle  \cdot \bs{n} \, \d s - \int_{S_{\bX^{(0)}}}\rho^{(0)} \left\langle \bv^{(1)}\otimes \bv^{(1)}\right\rangle  \cdot \bs{n} \, \d {s} =0. \label{ds_I}
\end{equation}
The linear problem (\ref{017})--(\ref{mbcI}) for $\av{\bv^{(2)}}$ and $\av{p^{(2)}}$ was solved explicitly by \citet{Doin1994R} with a vanishing right-hand side for (\ref{mbcI}); the difference is minor and the effect of the extra term is easy to track down. Introducing the result into (\ref{ds_I}) leads to the final equation for the particle motion
\begin{equation}
6\pi \eta a \left( \dot{\bX}^{(0)} - \tilde \bv \right) = \bs{F}_{\mathrm{ap}}, \label{meandyI}
\end{equation}
which has the form expected in our discussion of distinguished regimes in section \ref{strategy} (cf.\  (\ref{guessdyeq})). The velocity $\tilde \bv$  in (\ref{meandyI}) is given by
\begin{equation}
\tilde \bv  = \frac{1}{4\pi a^2}  \int_{S_{\bX^{(0)}}} \left( \av{\bv^{(2)}_\mathrm{i}}  + \av{X_r^{(1)} \frac{\partial \bv^{(1)}}{\partial r} } \right) \d s, \label{tildev}
\end{equation}
and  can be interpreted as a form of Lagrangian velocity averaged over the surface of the particle. Here, we have used axisymmetry to express the Stokes drift in terms of the $r$-component of the particle displacement $X_r^{(1)} = X^{(1)}\cos\theta$.
Note that while the second term in (\ref{tildev}) is the Stokes drift of fluid particles lying on the sphere, the first only includes the Eulerian-mean velocity associated with the incident part of the wave; as a result, $\tilde \bv$ differs from the full \added{averaged} Lagrangian velocity of these fluid particles.  \added{One feature of $\tilde \bv$ is its independence of the scattered Eulerian mean dynamics; another, discussed below, is its scaling for large and small viscosity.}

The acoustic pressure in (\ref{meandyI}) is given by
\begin{equation}
\bs{F}_{\mathrm{ap}} = - \frac{3 \pi \rho^{(0)}}{2} \sum_{n=0}^\infty \frac{n+1}{(2n+1)(2n+3)} \left(E_n A_n A^*_{n+1} + E_n^* A_n^* A_{n+1} \right) \,  \be_x. \label{forceI}
\end{equation} 
It depends on the wave amplitudes $A_n$ and on the coefficients $E_n$, which are slight modifications of  the $D_n$ computed by  \citet{Doin1994R} (his Eq.\ (5.6)). Specifically, our $E_n$ are deduced from the $D_n$ by setting the coefficients $S_{9n}$  to zero and omitting the terms proportional to the functions $G^{(l)}_n,\, L^{(l)}_n,\, K^{(l)}_n$  from the coefficients $S_{1n}$ to $S_{8n}$. This modification is made to include the Stokes-drift  term in the velocity $\tilde \bv$ whereas \citet{Doin1994R} includes it in his acoustic force (see also Ref.\ \citenum{Dani2000}). We find our choice  convenient for two reasons: (i)  all the terms in $\bs{F}_{\mathrm{ap}}$ depend on the scattered wave in the sense that $ \bs{F}_{\mathrm{ap}} \to 0$ as $\alpha_n, \, \beta_n \to 0$; and (ii) $\eta \tilde \bv$ and $\bs{F}_{\mathrm{ap}}$ have different behaviours in the limits of large and small viscosity. We discuss (ii) further in section \ref{LarVisReg}.

To summarise, the dynamics of the sphere in Regime I is controlled by a balance between a Stokes drag towards the streaming velocity $\tilde \bv$ and the acoustic pressure. We next illustrate the transient dynamics this leads to with the familiar example of a plane standing wave. 

\subsection{Standing wave}

For a plane standing wave, the potential of the incident wave can be expressed as
\begin{equation}
\phi_\mathrm{i} = A \cos(kr\cos\theta+kX^{(0)})\e^{-\ii\omega t}
 = \e^{-\ii\omega t} \sum \limits_{n=0}^{\infty} A_n j_n(kr) P_n(\cos\theta),
\end{equation}
where $A_n = \frac{1}{2}A(2n+1)i^n[\e^{\ii kX^{(0)}}+(-1)^n \e^{-\ii kX^{(0)}}]$. Note that the dependence of the $A_n$ on $X^{(0)}$ couples the wave field on the particle to the mean position $X^{(0)}$ of the particle. For simplicity, we consider the  particle motion in the particular case $\lambda=1$ and in the long wavelength limit $|ka| \ll 1, \, |k \delta| \ll 1$. 
In this limit, it can be shown using the asymptotics of Bessel functions \citep{dlmf2012} that the coefficients $E_n$ in (\ref{forceI}) satisfy $E_0=2 (ka)^3/9 \gg E_n, \, n \ge 1 $. As a result, the acoustic pressure  reduces to
\begin{equation}
\bF_{\mathrm{ap}} = -\frac{\pi \rho^{(0)} |A|^2 (k a)^3}{3} \sin(2k X^{(0)}) \, \be_x.
\end{equation}
Similarly, $\tilde \bv$ reduces to the Lagrangian-mean velocity of the incident wave (see section \ref{sphfol}), which vanishes for standing waves.  The position of the sphere therefore obeys the equation
\begin{equation}
\dot{X}^{(0)} = -\frac{\rho^{(0)} |A|^2 \sin(2k X^{(0)})(k a)^3}{18 \eta a},
\end{equation}
with solution
\begin{eqnarray}
X^{(0)} = k^{-1}\tan^{-1} \left( C \e^{-t/\tau}\right), \quad \textrm{where} \ \ 
\tau = \frac{9 \eta}{\rho^{(0)} |A|^2 k^4 a^2} \label{egI}
\end{eqnarray}
and $C$ is determined by the initial condition. This shows that the sphere converges exponentially towards the nodes of the standing waves over a time scale $\tau$. 

\section{Regime II}\label{RegII}

Regime II is characterised by $\alpha=1/2$, corresponding to a weaker dissipation than in Regime I, with $\delta/a = O(\eps^{1/2})$ rather than $O(1)$. As a result, the acoustic pressure on the particle is balanced by a combination of viscous drag and inertia, while the streaming velocity $\tilde \bv$ is negligible. The mean time scale is short compared to that in Regime I, $O(\eps^{-1})$ rather $O(\eps^{-2})$; crucially, this leads to mean velocities, both of the particle and of the surrounding fluid, that are comparable to the wave velocities.

These large mean velocities, and hence mean displacements necessitate to introduce coordinates that follow the motion of the particle. Defining $\br = \bx - \bX$, we rewrite the Navier--Stokes equations (\ref{012})--(\ref{015}) in these coordinates, noting  that $\nabla \mapsto \nabla_{\br}$ and $\partial_t \mapsto \partial_t - \dot \bX \cdot \nabla_{\br}$ to obtain
\begin{subequations} \label{NSmod}
\begin{align}
\frac{\partial }{\partial t}(\rho \bv) - \dot \bX  \cdot \nabla(\rho \bv) +\nabla \cdot ( \rho \bv\otimes\bv ) &= -\nabla p + \eta(\nabla^2\bv + \frac{1}{3}\nabla \nabla \cdot \bv) + \xi \nabla \nabla \cdot \bv, \label{NSmod1}\\
\frac{\partial \rho}{\partial t} -\dot \bX \cdot \nabla \rho + \nabla \cdot (\rho \bv)&= 0, \label{NSmod2}
\end{align}
\end{subequations}
where we have omitted the subscripts $\br$ from $\nabla_{\br}$ for convenience.

The weak viscosity of Regime II makes it possible to use a boundary-layer approach for both the wave and mean part of the dynamics. The boundary-layer thickness is $\delta=O(\eps^{1/2}a)$ so that all fields need to be expanded in powers of $\eps^{1/2}$ according to
\begin{subequations}\label{007}
\begin{align}
\bv&=\epsilon \bv^{(1)} +\epsilon^{3/2} \bv^{(3/2)} +\epsilon^2 \bv^{(2)} + \cdots, \\
\rho&= \rho^{(0)} + \epsilon \rho^{(1)} + \epsilon^{3/2} \rho^{(3/2)} + \epsilon^2 \rho^{(2)} + \cdots,\\
p&= p^{(0)} + \epsilon p^{(1)} + \epsilon^{3/2} p^{(3/2)} + \epsilon^2 p^{(2)} + \cdots,\\
\bX&= \bX^{(0)} + \epsilon \bX^{(1)} + \epsilon^{3/2} \bX^{(3/2)} + \epsilon^2 \bX^{(2)} + \cdots, 
\end{align}
\end{subequations} 
where $\rho^{(0)}$ and $p^{(0)}$ are constants. We anticipate that $\bX^{(0)}$ depends on the slow time $T= \eps t$ only, but all the other variables depend on both $t$ and $T$. We emphasise that $\bv^{(1)}$, $\rho^{(1)}$ and $p^{(1)}$ have both oscillatory and  mean contributions: we separate these two contributions using the notation
\[
\bv^{(1)} = \bar \bv^{(1)} + \bv'^{(1)}, \quad \textrm{with} \ \ \av{\bv'^{(1)}}=0.
\]

\subsection{Wave dynamics}

We now obtain the form of the leading-order wave fields. Substituting (\ref{007}) into (\ref{NSmod}) and the equation of state, and subtracting the mean contribution, we find the leading-order wave equations
\begin{subequations}\label{wgeII}
\begin{align}
\rho^{(0)} \frac{\partial {\bv'}^{(1)}}{\partial t} &= -\nabla p'^{(1)},\\
\frac{\partial \rho'^{(1)}}{\partial t} &= -\rho^{(0)}  \nabla \cdot \bv'^{(1)},\\
p'^{(1)} &= c^2 \rho'^{(1)}, 
\end{align}
\end{subequations}
with boundary conditions
\begin{subequations}\label{wbcII}
\begin{align}
\bv'^{(1)} &= \partial_t \bX'^{(1)} \quad  \mathrm{for} \ \ \br \in S_0, \label{wbcII1} \\ 
\bv'^{(1)} &\sim \bv_\mathrm{incident} \quad \mathrm{as} \ \ \br \to \infty, \label{wbcII2}
\end{align}
\end{subequations}
where $S_0$ denotes the sphere centred at origin. The equation for the sphere becomes
\begin{equation}
M\partial_{tt} \bX'^{(1)} = -\int\limits_{S_0}^{}p'^{(1)} \bn \, \d s. \label{wdeII}
\end{equation}
Eqs.\ (\ref{wgeII}) are the familiar equations for inviscid acoustic waves which can be solved in terms of a potential. 
As is standard, these equations are solved by imposing only the normal component of the boundary condition (\ref{wbcII1}). The potential solution so obtained is valid to leading order for $|r - a| \gg \delta$ only. Viscous effects are important in a boundary layer of thickness $\delta$   in which the velocity has a rotational contribution. The resulting  velocity tangential to the sphere varies rapidly so as to both match the potential solution and satisfy the no-slip condition. It turns out that the details of the solution in the boundary layer are unimportant for the leading-order dynamics of the sphere in Regime II; in particular, the effect of boundary streaming is $O(\eps^2)$ like that of interior streaming \citep{Nybo1965} and both contribute to the streaming velocity $\tilde \bv$ whose drag is $O(\eps^{5/2}\added{)}$ hence negligible.

The solution for $|r - a| \gg \delta$ is given, as in Regime I, by the sum $\phi_\mathrm{i}+\phi_\mathrm{s}$ of the incident and scattered wave, with 
\begin{equation}
\phi_\mathrm{i} = \e^{-\ii \omega t} \sum_{n=0}^\infty A_n j_n(k_0 r) P_n(\cos\theta).  \label{incexp}
\end{equation}
In this expression, the functions $A_n(\bX)$ which appear in the far-field condition (\ref{wbcII2})  when this is written in terms of $\br$ are approximated as $A_n(\bX^{(0)})$. The correction involving $\bX^{(1)} \cdot \nabla A(\bX^{(0)})$ is $O(\eps^2)$ and negligible. Note that the wavenumber can be taken as the inviscid approximation $k_0=\omega/c$, assuming implicitly that the far-field condition is imposed for some $\br$ not so large that the viscous decay (on scales given by $(\Im k)^{-1} \sim k_0^{-1} (k_0 \delta)^2 = O(k_0^{-1} \eps^{-1})$ for $k_0 \delta \ll 1$, see (\ref{k})) matter.  This damping introduces an outer scale that can modify the streaming, which is further discussed in \S \ref{Dis}. 
The potential of the scattered wave is given by
\begin{equation}
\phi_{\mathrm{s}} = \e^{-\ii\omega t} \sum_{n=0}^{\infty} \alpha_n A_n h_n(kr) P_n(\cos\theta),
\end{equation} 
where the coefficients are obtained from (\ref{wgeII})--(\ref{wdeII}) as
\begin{equation}
\begin{aligned}
\alpha_0 = -\frac{j_1(\kappa)}{h_1(\kappa)}, \quad
\alpha_1 = \frac{\lambda j_1(\kappa)-x j_1'(\kappa)}{\kappa h_1'(\kappa)-\lambda h_1(\kappa)}, \quad
\alpha_n = -\frac{j_n'(\kappa)}{h_n'(\kappa)} \ \ \textrm{for} \ \ n>1,
\end{aligned}
\end{equation} 
with $\kappa=k_0a$.
This result was first obtained by King \citep{King1934}.


\subsection{Mean dynamics} \label{MeanII}

We now turn to the mean dynamics. 
Time averaging the transformed Navier--Stokes equations (\ref{NSmod}), we obtain
\begin{subequations}\label{014}
\begin{align}
&\epsilon \av{ \frac{\mathbf{\partial {\rho\bv}}}{\partial T} }
-  \av{\frac{d \bX}{d t} \cdot \nabla (\rho \bv) } 
 + \nabla\cdot\av{\rho \bv \otimes \bv }  
\nonumber \\ 
& \qquad \quad = - \nabla {\bar{p}} + \eta \rb{\nabla^2\bar{\bv} + \frac{1}{3}\nabla(\nabla \cdot \bar{\bv})} + \xi \nabla(\nabla \cdot \bar{\bv}), \label{mme} \\
&\epsilon \frac{\partial \bar{\rho}}{\partial T} -  \av{\frac{d \bX}{d t} \cdot \nabla \rho}  + \nabla \cdot \langle \rho \bv \rangle = 0, \label{mmce}
\end{align}
\end{subequations}
where $d/dt = \partial_t + \eps \partial_T$. The boundary condition is given by
\begin{equation}
{\bar{\bv}} = \dot{\bX}^{(0)} \, \, \mathrm{for} \, \, \boldsymbol{r}\in S_0, \label{027}
\end{equation}
and an prescribed outer boundary condition depending on the specific application.

Corresponding to the wave solution, the mean flow has a boundary layer of thickness $\delta$ around the particle. We therefore analyse the mean equations (\ref{014}) separately in an outer region with $r-a \gg \delta$ and in a boundary layer with $r-a = O(\delta)$.  

\subsubsection{Outer region}\label{OutReg}
Substituting (\ref{007}) into the mean mass-conservation equation (\ref{mmce}) gives
\begin{subequations}
\begin{align}
&O(\epsilon): \quad \quad \nabla \cdot \bar{\bv}^{(1)} = 0, \label{023}\\
&O(\epsilon^{3/2}): \quad \nabla \cdot \bar{\bv}^{(3/2)} = 0, 
\end{align}
\end{subequations}
which imply that both $\bar{\bv}^{(1)}$ and $\bar{\bv}^{(3/2)}$ are incompressible.

Similarly, the mean momentum equation (\ref{mme}) gives
\begin{subequations}
\begin{align}
&O(\epsilon): \quad 0 = - \nabla \bar{p}^{(1)},\\
&O(\epsilon^{3/2}): \quad 0 = - \nabla \bar{p}^{(3/2)},\\
\begin{split}
&O(\epsilon^2): \quad \rho^{(0)} \frac{\partial \bar{\bv}^{(1)}}{\partial T} 
- \rho^{(0)}  \dot{\bX}^{(0)}\cdot \nabla \bar{\bv}^{(1)}
-\rho^{(0)}  \av{\frac{\partial \bX'^{(1)}}{\partial t} \cdot \nabla {\bv'}^{(1)} } \\
& \qquad \qquad + \nabla \cdot  \av{\rho^{(0)} {\bv'}^{(1)} \otimes {\bv'}^{(1)} }
+ \nabla \cdot  \rb{\rho^{(0)} \bar{\bv}^{(1)} \otimes \bar{\bv}^{(1)}} 
= -\nabla \bar{p}^{(2)} + \hat{\eta} \nabla^2 \bar{\bv}^{(1)}, \label{031}
\end{split}
\end{align}
\end{subequations}
where (\ref{023}) is used and we have defined $\hat \eta = \eta/\eps = O(1)$ consistent with the assumption that $\delta=O(\eps^{1/2})$.
Assuming that $\bar{p}^{(1)}$ and $\bar{p}^{(3/2)}$ tend to constants as $|\boldsymbol{r}|\to\infty$, we conclude that $\bar{p}^{(1)}$ and $\bar{p}^{(3/2)}$ are constant in the outer region. The key equation is (\ref{031}) which describes the mean dynamics in the outer region; its boundary conditions are obtained by considering the boundary layer.

\subsubsection{Boundary layer}

This region is defined by $R = (r-a)/\delta = O(1)$. Denoting the dependent variables regarded as functions of $R$ and $\theta$ by capital letters, we obtain from the
mass conservation (\ref{mmce}) that
\begin{subequations}\label{messcon}
\begin{align}
&O(\epsilon^{1/2}): \quad \rho^{(0)} \frac{\partial \bar{V}^{(1)}_r}{\partial R} = 0, \label{028}\\
&O(\epsilon): \quad 2\bar{V}^{(1)}_r + \frac{1}{\sin\theta}\frac{\partial}{\partial \theta} (\sin\theta \, \bar{V}^{(1)}_\theta) + \frac{\partial \bar{V}^{(3/2)}_r}{\partial R} = 0. \label{029}
\end{align}
\end{subequations}
Eq.\ (\ref{028}) indicates that $\bar{V}_r^{(1)}$ is independent of $R$ across the boundary layer: $\bar{V}^{(1)}_r(R,\theta) = \bar{v_r}^{(1)}(r=a,\theta)$.
The mean momentum conservation (\ref{mme}) gives
\begin{subequations}
\begin{align}
O(\epsilon^{1/2}): \quad & 0 = -\frac{\partial \bar{P}^{(1)}}{\partial R},\label{p1} \\ 
O(\epsilon): \quad & 0 = -\frac{\partial \bar{P}^{(3/2)}}{\partial R} , \label{p2}\\
				& 0 = -\frac{1}{a}\frac{\partial \bar{P}^{(1)}}{\partial \theta} + \frac{\hat{\eta}}{a^2} \frac{ \partial^2 \bar{V}^{(1)}_\theta }{\partial R^2}, \label{MMtheta}
\end{align}
\end{subequations}
when (\ref{028}) is used. We conclude from (\ref{p1}) and (\ref{p2}) that $\bar{P}^{(1)}$ and $\bar{P}^{(3/2)}$ are constant across the boundary layer, so that $\bar{p}^{(1)}$ and $\bar{p}^{(3/2)}$ are constant throughout the fluid. It then follows from (\ref{MMtheta}) that $\partial^2 \bar{V}^{(1)}_\theta/\partial R^2 = 0$, hence
\begin{equation}
\bar{V}^{(1)}_\theta = f_1(\theta) R + f_2(\theta),
\end{equation}
where the functions $f_1$ and $f_2$ remain to be determined. Matching with the outer solution gives that $f_1 = a\partial_r \bar{\bv}^{(1/2)}(r=a,\theta) = 0$ (since $\bar{\bv}=O(\eps)$), and $f_2=\bar{v}^{(1)}_\theta(r=a,\theta)$. This implies that $\bar{\boldsymbol{V}}^{(1)}$ is independent of $R$ across the boundary layer. As a result, the outer velocity satisfies the simple boundary condition 
\begin{equation} \label{vVX}
{\bar{\bv}}^{(1)}(r=a,\theta)= {\bar{\boldsymbol{V}}}^{(1)}(r=a,\theta) = \dot{\bX}^{(0)}.
\end{equation}

The momentum equation obtained at the next order, $O(\eps^{3/2})$, can be reduced using (\ref{028}) and (\ref{029}) to
\begin{subequations}\label{II3/2}
\begin{align}
&\frac{\rho^{(0)}}{a} \av{(V'^{(1)\phi}_r - \frac{\partial X'^{(1)}}{\partial t}\cos\theta ) \frac{\partial V'^{(1)\psi}_\theta}{\partial R}}  + \frac{\rho^{(0)}}{a} \rb{\bar{V}^{(1)}_r-\dot{{X}}^{(0)}\cos\theta} \frac{\partial \bar{V}^{(1)}_\theta}{\partial R} \nonumber \\
&= - \frac{1}{a}\frac{\partial \bar{P}^{(3/2)}}{\partial \theta} + \frac{\hat{\eta}}{a^2} \frac{\partial^2 \bar{V}^{(3/2)}_\theta}{\partial R^2} +  \frac{2 \hat{\eta}}{a^2} \frac{\partial \bar{V}^{(1)}_\theta}{\partial R},\\
\quad \quad 0 &= - \frac{1}{a}\frac{\partial \bar{P}^{(2)}}{\partial R} + \frac{\hat{\eta}}{a^2} \frac{\partial^2 \bar{V}^{(3/2)}_r}{\partial R^2} +  \frac{2 \hat{\eta}}{a^2} \frac{\partial \bar{V}^{(1)}_r}{\partial R},
\end{align}
\end{subequations}
where the mass conservation (\ref{messcon}) and wave solutions have been used. Here $V'^{(1)\phi}_r$ and 
$V'^{(1)\psi}_\theta$ denote potential and rotational contributions to the wave velocity $\boldsymbol{V}'^{(1)}$. Since, as discussed above, the potential part satisfies the no-normal flow condition, the term involving these contributions vanishes. Using the constancy of $P^{(3/2)}$ and 
(\ref{vVX}) reduces (\ref{II3/2}) to
\begin{subequations}
\begin{align}
0 &=  \hat{\eta} \frac{\partial^2 \bar{V}^{(3/2)}_\theta}{\partial R^2}, \label{029.5} \\
0 &= - \frac{1}{a} \frac{\partial \bar{P}^{(2)}}{\partial R} + \frac{\hat{\eta}}{a^2} \frac{\partial^2 \bar{V}^{(3/2)}_r}{\partial R^2}. \label{030}
\end{align}
\end{subequations}
Therefore, $\bar{V}^{(3/2)}_\theta = f_3(\theta) R + f_4(\theta)$, where $f_3$ and $f_4 $ are obtained by matching as $f_3 = a\partial_r \bar{v}^{(1)}_{\theta}(r=a,\theta) $  and $f_4=\bar{v}^{(3/2)}_\theta(r=a,\theta)$.
Taking the $R$ derivative of (\ref{029}) yields $\partial^2 \bar{V}^{(3/2)}_r/\partial R^2 = 0$, which implies $\bar{V}^{(3/2)}_r = a \partial_r \bar{v}^{(1)}_r(r=a,\theta) R + \bar{v}^{(3/2)}_r(r=a,\theta)$ after matching. Therefore, 
(\ref{030}) reduces to $\partial \bar{P}^{(2)}/\partial R = 0$ so that $\bar{p}^{(2)}$ is $R$-independent in the boundary layer.

The above calculation provides us with two important pieces of information: (i) the $O(\epsilon)$ velocity and $O(\epsilon^2)$ pressure are $R$-independent; and (ii) the $O(\epsilon^{3/2})$ velocity depends linearly on $R$. From this, we conclude that the stress is constant across the boundary layer up to $o(\eps^2)$ corrections. As a result, the leading-order particle motion, which depends only on the $O(\eps^2)$ stress,  can be computed from the outer solution alone.

\subsubsection{Governing equations} 

The previous two sections conclude that the mean dynamics is controlled by the incompressible-fluid  momentum equation (\ref{031}) and the mean particle  equation
\begin{equation}
M \ddot{\bX}^{(0)} = -\int\limits_{S_0}^{} \bar{p}^{(2)} \bs{n} \, \d s + \int\limits_{S_0}^{} \bar{\tau}^{(2)} \cdot \bs{n} \, \d s, \label{IISph}
\end{equation}
that arises when (\ref{007}) is introduced into (\ref{newton}). These two equations are coupled through  
 the mean stress tensor $\bar{\tau}$, defined by
 \begin{equation}
\begin{aligned}
\bar{\tau}^{(2)}_{rr} &= -2 \hat{\eta }\frac{\partial \bar{v}^{(1)}_r}{\partial r}, \\
\bar{\tau}^{(2)}_{\theta\theta} &= -2 \hat{\eta} \left( \frac{1}{r} \frac{\partial \bar{v}^{(1)}_\theta}{\partial \theta} + \frac{\bar{v}^{(1)}_r}{r} \right), \\
\bar{\tau}^{(2)}_{r\theta} = \bar{\tau}^{(2)}_{\theta r} &= -\hat{\eta} \left(  r\frac{\partial}{\partial r}\left( \frac{\bar{v}^{(1)}_\theta}{r} \right) +\frac{1}{r}\frac{\partial \bar{v}^{(1)}_r}{\partial \theta} \right),
\end{aligned}
\end{equation}
and through the no-slip  condition (\ref{vVX}) satisfied by $\bar{\bv}^{(1)}$. We now recast these equations in a simpler form and discuss the physical mechanism they describe.

The effect of the waves on the particle is implicit in (\ref{IISph}): it arises through changes in  $\bar{p}^{(2)}$ and $\bar \bv^{(1)}$ that are induced by the presence of  wave terms in the momentum equation (\ref{031}). We can make the effect of the waves explicit in (\ref{IISph}) by writing these terms using the wave potential in the outer region\citep{Gork1962} to obtain
\begin{subequations}
\begin{align}
\nabla \cdot\left\langle \bv'^{(1)}\otimes \bv'^{(1)}\right\rangle &= \nabla  \left\langle \frac{1}{2}\left(\nabla \phi'^{(1)}\right)^2 - \frac{1}{2c^2}\left(\frac{\partial \phi'^{(1)}}{\partial t}\right)^2 \right\rangle,\\
\av{\frac{\partial \bX'^{(1)}}{\partial t} \cdot \nabla \mathbf{v'}^{(1)} }
&= \nabla\left\langle  \frac{\partial \bX'^{(1)}}{\partial t} \cdot \nabla \phi'^{(1)} \right\rangle.
\end{align}
\end{subequations}
It is therefore natural to redefine pressure as
\begin{equation}
\tilde{p}^{(2)} = \bar{p}^{(2)} + \rho^{(0)} \left\langle \frac{1}{2}\left(\nabla \phi'^{(1)}\right)^2 - \frac{1}{2c^2}\left(\frac{\partial \phi'^{(1)}}{\partial t}\right)^2 \right\rangle  - \rho^{(0)} \left\langle \frac{\partial \bX'^{(1)}}{\partial t} \cdot \nabla \phi'^{(1)} \right\rangle,
\end{equation}
leading to the simpler momentum equation
\begin{equation}
\begin{aligned}
\rho^{(0)} \frac{\partial \bar{\bv}^{(1)}}{\partial T} 
- \rho^{(0)} \dot{\bX}^{(0)}\cdot \nabla \bar{\bv}^{(1)}
+  \rho^{(0)} \nabla \cdot   \rb{\bar{\bv}^{(1)} \otimes \bar{\bv}^{(1)}} 
= -\nabla \tilde{p}^{(2)} + \hat{\eta} \nabla^2 \bar{\bv}^{(1)}. \label{032}
\end{aligned}
\end{equation}
The advection term $\rho^{(0)}  \dot{\bX}^{(0)} \cdot \nabla \bar{\bv}^{(1)}$ in the above equation can be eliminated by using the  spatial coordinates
\begin{equation}
\tilde{\bx} = \bs{r} + \bX^{(0)} = \bx + O(\eps), \label{bcII}
\end{equation}
which can be identified with the original, fixed-frame coordinates $\bx$, as the second equality indicates.
This reduces (\ref{032}) to
\begin{equation}
\begin{aligned}
\rho^{(0)} \frac{\partial \bar{\bv}^{(1)}}{\partial T} 
+  \rho^{(0)} \nabla \cdot  \rb{\bar{\bv}^{(1)} \otimes \bar{\bv}^{(1)}} 
= -\nabla \tilde{p}^{(2)} + \hat{\eta} \nabla^2 \bar{\bv}^{(1)}, \label{033}
\end{aligned}
\end{equation}
where the spatial derivatives are with respect to $\tilde{\bx}$. Eq.\ (\ref{033}), together with the incompressibility condition 
\begin{eqnarray}
\nabla\cdot\bar{\bv}^{(1)}=0 \label{Incom},
\end{eqnarray} 
are the usual incompressible Navier--Stokes equations.
At the same time, equation (\ref{IISph}) for the sphere becomes
\begin{equation}
M \ddot{\bX}^{(0)} = -\int\limits_{S_{\bX^{(0)}}}^{} \tilde{p}^{(2)} \bs{n} \, \d s + \int\limits_{S_{\bX^{(0)}}}^{} \bar{\tau} \cdot \bs{n} \, \d s + \bs{F}_{\mathrm{inv}}, \label{004}
\end{equation}
where
\begin{equation}
\bs{F}_{\mathrm{inv}} =  \rho^{(0)} \int\limits_{S_{\bX^{(0)}}}^{}  \left\{ \left\langle \frac{1}{2}\left( \nabla \phi'^{(1)}\right)^2 - \frac{1}{2c^2}\left(\frac{\partial \phi'^{(1)}}{\partial t}\right)^2 \right\rangle  -  \left\langle \frac{\partial \bX'^{(1)}}{\partial t} \cdot \nabla \phi'^{(1)} \right\rangle \right\}  \bs{n} \, \d s \label{ivF}
\end{equation}
is the inviscid acoustic pressure. The boundary conditions for (\ref{033}) become
\begin{equation}
\bar{\bv}^{(1)}= \dot{\bX}^{(0)} \quad \textrm{for} \ \ \tilde \bx \in S_{\bs{X}^{(0)}}, \label{NonSli}
\end{equation}
together with a condition on the outer of the fluid region, at infinity for instance. The latter boundary condition is naturally expressed in terms of the Lagrangian-mean velocity which, in particular, vanishes on the surface of oscillating wavemakers \cite{Brad1996}. This velocity can be identified with $\bar{\bv}^{(1)}$, however, since the Stokes drift is $O(\eps^2)$  hence negligible.

The force in (\ref{ivF}) is that obtained for a purely inviscid fluid and used, e.g., by \citet{Gork1962}. Explicit expressions for this force in terms of the coefficients $A_n$ in the expansion (\ref{incexp}) of the incident wave are given in Refs.\ \citenum{King1934,Doin1994R} and, in the long-wave limit $k a \ll 1$, in Ref.\ \citenum{Gork1962}.
Note that Gor'kov uses an integration over a large sphere rather over the particle itself, replacing $\partial_t \bX'^{(1)}$ by $\nabla \phi'^{(1)}$ and taking advantage of the divergence-free property of the mean inviscid stress tensor (the integrand). In our case, because the mean flow is affected by viscosity and acts on the particle through the first two terms on the right-hand side of (\ref{004}), this technique is not as useful.

To summarise, in Regime II, the slow, averaged dynamics is controlled by the coupled system (\ref{033})--(\ref{NonSli}).
In this system, viscosity enters only in the Navier--Stokes equation governing the fluid motion and not in the acoustic pressure.
This is a complex, nonlinear system involving a moving boundary, but a classical one, describing the motion of an
externally forced spherical particle in incompressible viscous fluid. It has been studied extensively, both theoretically (e.g., to find approximate solutions \citep{Bass1888,Maxe1983}) and numerically \citep{Jens1959,Rimo1969,Denn1972}.
The specificity of our problem is the form of the external force, namely the inviscid acoustic pressure which can be obtained solely from the potential wave solution.

\subsection{Example: plane standing wave} \label{IIeg}
To illustrate the difference between Regimes I and II, 
we consider again a standing wave in the limit $|ka|\ll 1$, $|k\delta|\ll 1$ for $\lambda=1$.
The acoustic pressure is then
\begin{equation}
\bF_{\mathrm{inv}} = -\frac{\pi \rho^{(0)} |A|^2 (k a)^3}{3} \sin(2k X^{(0)}) \, \be_x \label{eunap}
\end{equation}
and identical to that of Regime I \cite[e.g.][]{King1934,Doin1994R}.
To obtain a simple closed-form solution, we restrict our attention to the Basset limit \cite{Bass1888} where $\dot \bX^{(0)}$ is small enough that the advection terms can be treated perturbatively. This reduces (\ref{033})--(\ref{NonSli}) to the single equation
\begin{equation}
\begin{aligned}
(M+M') \ddot{X}^{(0)} = -DX^{(0)} -K \dot{X}^{(0)} - B\int\limits_{0}^{T}\ddot{X}^{(0)}(T-\tau) \, \tau^{-1/2} d\tau , \label{024}
\end{aligned}
\end{equation}
where we have linearised the acoustic pressure (\ref{eunap}) using that $X^{(0)} \ll 1$. The coefficients 
\begin{equation}
M' = \frac{2\pi}{3} a^3\rho^{(0)}, ~ 
D = \frac{2}{3}\pi \rho^{(0)} |A|^2 k_0^4 a^3, ~
K = 6\pi \hat\eta a, ~
B = 6 a^2(\pi \rho^{(0)} \hat\eta)^{1/2}
\end{equation}
can be associated with distinct physical effects: added mass, acoustic pressure, Stokes drag, and the (history-dependent) Basset force.

\begin{figure}
\begin{center}
\includegraphics[width=12cm]{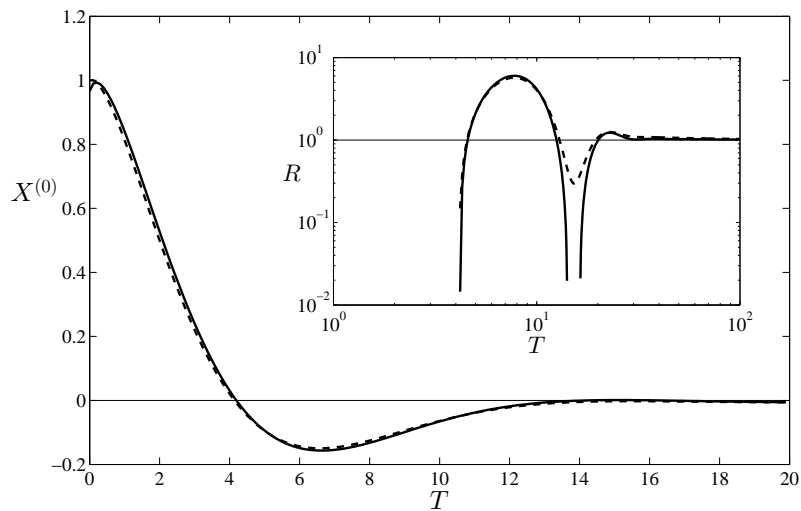}
\caption{Particle position $X^{(0)}$ as a function of $T$ for a standing wave in the Basset limit of Regime II obtained by solving (\ref{024}) analytically (solid line, see Appendix \ref{appf}) and numerically (dashed line). The long-time asymptotics is confirmed in the log-log coordinate inset which displays $R$, the ratio of the exact solution to (\ref{025}). The parameters are chosen as $M=1$, $B=1$, $K=1$ and $D=1$.}
\label{fig:2+1modes}
\end{center}
\end{figure}

We solve (\ref{024}) with $X^{(0)}(0)=X$ and $\dot X^{(0)} = 0$ using Laplace transform (see Appendix \ref{appf} for details).
The solution is the sum of exponentially damped oscillations and a continuous-spectrum contribution. This controls the $T \gg 1$ asymptotics, given by
\begin{equation}
X^{(0)} \sim 
 -\frac{BX}{2D} T^{-3/2}. \label{025}
\end{equation}
As in  Regime I, the particle tends to its equilibrium position at the node $X^{(0)}(0)=0$ of the standing wave. The differences are that the process is not monotonic, with the particle oscillating around the node, and is much slower than in Regime I, with a $T^{-3/2}$ decay of the distance to the node rather than the exponential decay of Regime I. This is illustrated in Figure \ref{fig:2+1modes}.

\added{
\subsection{An exceptional case: plane travelling wave}

Plane travelling waves are an important exception to the generic scaling of Regime II because, for small viscosity, the acoustic pressure they exert turns out to be much smaller than the $O(\eps^2)$ assumed so far. Specifically, it is smaller by a factor $O(\delta/a) = O(\epsilon^{\alpha})$ when $\delta/a \gg (ka)^3$, an assumption we make here  (see Eq.\ (6.13) in Ref.\ \citenum{Doin1994R}). The dominant-balance argument of section \ref{regimes} needs to be revisited for this case. It is easily checked that, with an acoustic pressure of relative order $\eps^{2+\alpha}$ rather than $\eps^2$, the balance between acceleration, drag and acoustic pressure that characterises Regime II is obtained for $\alpha=2/3$ and $\beta=4/3$. As a result, the mean flow $\bar{\bs{v}}$ is of order $O(\epsilon^{4/3}c)$, asymptotically smaller than the $O(\epsilon c)$ wave velocity $\bs{v}'$.
The wave-mean flow interaction is much weaker than in the generic case above: because the drag exerted by the mean flow on the sphere is negligible, there is no need to solve the full Navier-Stokes equations, and dynamics of the sphere is governed by the simple equation
\begin{equation}
M \ddot{\bX}^{(0)}+6\pi\eta a \dot{\bX}^{(0)} = \bF_\mathrm{ap}^{\mathrm{t}}.
\end{equation} 
The explicit form of $\bF_\mathrm{ap}^{\mathrm{t}}$, where  the superscript $\mathrm{t}$ highlights the applicability to plane travelling waves only, can be found in Ref.\ \citenum{Doin1994R} as Eq.\ (6.13).
}

\section{Intermediate regimes}\label{IntReg}

\begin{table}
\begin{center}
    \begin{tabular}{ l  c  c }
    \hline
    Regime & Parameter range & $\alpha$ \\ \hline \hline
    Large-viscosity regime & $\delta/a \gg 1$ & $\alpha<0$\\ 
        Regime I & $\delta/a = O(1)$ & $\alpha=0$ \\ 
    Transition regime & $\epsilon^{1/2} \ll \delta/a \ll 1$ & $0 < \alpha < 1/2$ \\ 
    Regime II & $\delta/a = O(\epsilon^{1/2})$ & $\alpha=1/2$ \\ 
    Inviscid regime & $\delta/a \ll \epsilon^{1/2}$ & $\alpha > 1/2$ \\ \hline
    \end{tabular}
    \caption{The asymptotic regimes considered in this paper are characterised by  relations between $\delta/a$ and $\eps \ll  1$ or, equivalently, by $\alpha$ such that $\delta/a = \eps^\alpha$.} \label{RegTab}
    \end{center}
\end{table}

Regimes I and II are distinguished regimes characterised by specific scaling relations between the parameters $\delta/a$ and $\epsilon$. We now briefly consider intermediate regimes which can be regarded as sublimits of Regimes I and II; these apply over broad regions of the $(\delta/a,\eps)$ space and yield much simplified governing equations. 
These intermediate regimes  are listed in Table \ref{RegTab} together with the asymptotic inequalities that define them. A particularly important regime from the theoretical viewpoint is defined by $\epsilon^{1/2} \ll \delta/a \ll 1$ which marks the transition between Regimes I and II. By showing  that the mean equation in this transition regime is the limit of those in both 
 Regime I and Regime II, we confirm that our heuristic arguments in section\,2 identify all possible distinguished regimes. 

\subsection{Large-viscosity regime} \label{LarVisReg}

This regime corresponds to a large viscosity and is deduced from Regime I by letting $\eta \to \infty$.
It can be shown that the coefficients $E_n$ in the acoustic pressure (\ref{forceI}) remain bounded in this limit so $\bs{F}_\mathrm{ap}$ is negligible. This motivated our separation between $6 \pi \eta \tilde \bv$ and $\bs{F}_\mathrm{ap}$ in (\ref{forceI}).  Physically, these terms describe two very  different effects. The first is a linear (Stokes) drag controlled by the moving boundary; the second is controlled by the  average wave momentum flux and pressure which are bounded as $\eta \to \infty$.
As a result,  (\ref{meandyI}) reduces to
\begin{eqnarray}
\dot{\bX}^{(0)}= {\tilde{\bv}}. \label{largevis}
\end{eqnarray}
This simple balance is important for practical purpose since it allows for the possibility of particles following fluid elements as is required when using tracer particles; we this discuss this point further in section\,\ref{sphfol}. 

\subsection{Transition regime}

We  now show that the Regimes I and II overlap: specifically, the small-viscosity limit of Regime I matches the large-viscosity limit of Regime II in a transition regime where both streaming velocity and particle acceleration are negligible.

Starting from Regime I, we let $\eta \to 0$ in (\ref{meandyI}): the acoustic pressure then reduces to its inviscid form, while the viscous drag term $\eta \tilde \bv$ becomes negligible, leading to the balance
\begin{equation}
6\pi \eta a\dot{\bX}^{(0)} = \bs{F}_{\mathrm{inv}} \label{TranDy}.
\end{equation}

Conversely, letting $\hat \eta \to \infty$ in the mean momentum equation of Regime II, (\ref{033}), reduces this to the Stokes equation 
\begin{equation}
\begin{aligned}
0 = -\nabla \tilde{p}^{(2)} + \hat{\eta} \nabla^2 \bar{\bv}^{(1)},
\end{aligned}
\end{equation}
Since $\nabla \cdot \bar{\bv}^{(1)} = 0$, $\bar \bv^{(1)}$ is a Stokes flow around the spherical particle and the associated stress (first two terms on the right-hand side of (\ref{004})) is the familiar linear Stokes drag. Since, furthermore, the particle's acceleration is negligible, (\ref{TranDy}) is recovered.

\subsection{Inviscid regime}


We now consider the limit where viscosity is so small as to be negligible in both the acoustic pressure and flow equation. Letting  $\hat \eta \to 0$ in (\ref{033}) yields the Euler equation
\begin{equation}
\rho^{(0)}\frac{\partial \bar{\bv}^{(1)}}{\partial T}+\rho^{(0)}\bar{\bv}^{(1)}\cdot \nabla \bar{\bv}^{(1)} = -\nabla \tilde{p}^{(2)}.
\end{equation}
Assuming a potential mean flow $\bar{\bv}^{(1)}=\nabla \bar{\phi}^{(1)}$,
the pressure is expressed as 
\begin{equation}
\tilde{p}^{(2)} = \rho^{(0)}\frac{\partial \bar{\phi}^{(1)}}{\partial T} +\rho^{(0)}\frac{|\nabla \bar{\phi}^{(1)}|^2}{2}.
\end{equation}
The force on the particle associated with this pressure (first term on the right-hand side of (\ref{004})) is the well-known added-mass effect (e.g. Ref. \citenum{Batc1967}). This reduces (\ref{004}) to
\begin{equation}
(M+M') \ddot{{\bX}}^{(0)} = \bs{F}_{\mathrm{inv}}, \label{SmaVis}
\end{equation} 
where the added mass $M'=2\pi a^3 \rho^{(0)}/3$. This provides a consistent derivation of the added-mass effect incorporated by \citet{King1934} in an \textit{ad hoc} manner. 


\subsection{When do spherical particles follow fluid elements?} \label{sphfol}

Rigid spherical particles are often used in experiments as passive tracers to visualise and quantify fluid motion. It is therefore important to find conditions that ensure the spherical particles follow closely the motion of the fluid elements they are  meant to trace, without being disturbed by the scattering induced by the rigid particle itself. In the presence of acoustic waves, fluid elements move (on average) with the Lagrangian mean velocity, so the conditions should ensure that $\dot{\bX}^{(0)} = \av{ \bv_\mathrm{i}^{(2)} +  \bs{\xi}^{(1)}_\mathrm{i} \cdot \nabla \bv_\mathrm{i}^{(1)} }$ up to negligible errors. Here $\bs{\xi}^{(1)}_\mathrm{i}$ is the displacement of fluid elements associated with the incident wave only. We now show that sufficient conditions for this are that
\begin{equation}
\delta/a \gg 1, \ |ka| \ll 1 \ \ \textrm{and} \ \ \lambda = 1. \label{sfp}
\end{equation}

The first condition places the dynamics in the large-viscosity regime discussed in section \ref{LarVisReg} in which the acoustic pressure is negligible so that $\mathbf{\dot{\bX}^{(0)}}=\tilde{\bv}$. The second condition ensures that the average over the sphere in (\ref{tildev}) is a good approximation to the value of the integrand at the centre of the sphere. It remains to show that $X_r^{(1)} \partial_r \bv^{(1)} \approx \xi_{\mathrm{i}r}^{(1)} \partial_r \bv^{(1)}_\mathrm{i}$ to ensure that the second term in (\ref{tildev}) approximates the Stokes drift associated with the incident wave.  

To show this, we consider the leading-order particle velocity
\begin{equation}
u=\frac{\lambda A_1 \bs{e}_x}{ka}\left[j_1(ka)+\alpha_1 h_1(ka) + 2\beta_1 h_1((1+\ii)\delta/a)\right] \e^{-i\omega t},
\end{equation}
with coefficients $\alpha_1$ and $\beta_1$ given explicitly by \citet{Doin1994J}. It can be checked that
$\lambda = 1$ leads to $\beta_1=0$ and, using the asymptotics of spherical Bessel functions \citep[e.g.,][]{dlmf2012}, that $|ka| \ll 1$ leads to $\alpha_1 = o(|ka|)$. Therefore the velocity of the sphere reduces to
$u = A_1 \bs{e}_x \e^{-\ii\omega t} + o(|ka|A_1)$. To leading order, this coincides with the radial velocity of the incident wave: indeed, for $|ka| \ll 1$ thisvelocity  is dominated by the mode $n=1$ in the expansion (\ref{incexp0}) while the scattered wave is negligible, again because $\alpha_1 = o(|ka|)$ and $\beta_1=0$. It follows that $X_r^{(1)} =\xi_{\mathrm{i}r}$. Similarly, the wave velocity around the sphere is dominated by the incident component, hence $\partial_r \bv^{(1)} = \partial_r \bv^{(1)}_\mathrm{i}$.

\section{Discussion}\label{Dis}

In this paper, we examine the dynamics of a spherical particle in an axisymmetric acoustic field and derive simplified models governing the mean motion of the particle. This is controlled by the complex interaction of the particle with both the wave and the surrounding fluid. Specifically, four physical effects come into play: inertia (of the particle and of the fluid it entrains), viscous drag, acoustic streaming, and acoustic pressure. Under the assumption $k a = O(1)$, or more accurately $ka=O(\eps^\gamma)$ for $\gamma \ge 0$, of a particle that is of the same order as or smaller than the acoustic wavelength, our analysis shows that these four effects are never concurrent. Depending on the strength of viscosity (measured by the parameter $\delta/a$) relative to the wave amplitude, several regimes, characterised by the balance between two or three of these effects, are possible. These are listed in Table \ref{RegTab} and we briefly summarise their main features below.

Fixing the radius of the particle, the relevant regime is determined by the value of viscosity. For large viscosity, the 
particle is driven by a viscous response to the streaming velocity, with negligible acoustic pressure and inertia, leading to (\ref{largevis}). As viscosity decreases, the drag effect decreases and acoustic pressure becomes significant, leading to the three-term balance between drag, streaming and acoustic pressure of Regime I and the more complex model (\ref{meandyI}) for $\delta/a = O(1)$. For smaller viscosity still, the streaming effect becomes negligible so that viscous drag balances acoustic pressure, yielding Eq.\ (\ref{TranDy}). Importantly, in this
`transition regime'  viscosity is weak enough for the acoustic pressure to be well approximated by its inviscid form.  When  viscosity is such that $\delta/a=O(\eps^{1/2})$, particle inertia comes into play. This is  Regime II, where inertia, fluid stress (associated with both pressure and viscosity) and (inviscid) acoustic pressure balance. This is a rather complex regime in which the mean dynamics of the particle and of the fluid are fully coupled and the Navier--Stokes equations need to be solved to determine the fluid stress acting on the particle. The mean equations of motion are then (\ref{033})--(\ref{NonSli}). Finally, for very weak viscosity, the fluid motion is governed by the Euler equation and, under the assumption of a potential flow, its impact reduces to the familiar added-mass effect, leading to (\ref{SmaVis}). Regimes I and II are of particular importance because they correspond to distinguished, three-term balances, and encompass the other regimes as specific sublimits.

It is interesting to note that various acoustic microfluidic experiments span a range of parameters and hence a range of regimes. We have estimated the parameters used in several experiments that employ particles for a variety of purposes in order to assess which dynamical regime is relevant to each. The experiments of \citet{From2008} use spherical particles to trace the mixing flow generated by time-dependent acoustic streaming. The key non-dimensional parameters are approximately 
$\delta/a \approx 0.13$ and $\epsilon^{1/2} \approx 0.03$. This places these experiments in the transition regime and indicates that  acoustic pressure could affect the particle and cause their trajectories to depart from those of fluid elements. However, for these experiments, the streaming velocity is substantially larger than our estimate, $O(\eps^2)$ non-dimensionally or ${v'}^2/c$ dimensionally: as discussed in Ref.\ \citenum{Vann2011} (for $k \delta \ll 1)$  in problems where the wave amplitudes vary over an outer scale $\ell$ that differs from $k$ (e.g., for weakly damped travelling wave), the streaming velocity is $O((k\ell)^2\eps^2)$. In such cases, the viscous drag can dominate the acoustic pressure even though $\delta/a$ is not large. In the particle collection experiments of \citet{Li2007}, \citet{Ober2009} and  \citet{Tan2009}, the values of $\delta/a$ are $O(1)$ (1.28, 0.19--0.33 and 0.52--0.64, respectively), placing the experiments in Regime I. Interestingly, 
\citet{Li2007} observe  concentration times proportional to $a^{-2}$, consistent with (\ref{egI}). In another set of particle collection experiments, \citet{Roge2010} use a broader range of particle diameter, such that $\delta/a \in [0.017, 0.9]$. Estimating their wave amplitudes be in the range $\epsilon \in [9.5\times 10^{-5},2.9\times 10^{-4}]$, we conclude that the experiments span both Regimes I and  II.

While this paper concentrates on spherical particles that are of the same order as or smaller than the wavelength, we can sketch how the analysis could be extended to larger particles, with $k a = O(\eps^\gamma)$ for $\gamma < 0$. 
The main difference for the balance of terms in the equation governing the particle motion is the order of magnitude of the acoustic pressure. Recall that in the heuristic model (\ref{guessdyeq}), this was taken to be $O(\eps^2)$, corresponding to the dimensional estimate $k a^3 {v'}^2$ and to the assumption that the length scale for the change of momentum flux over the sphere is proportional to $k^{-1}$. For $\gamma < 0$, this scale is instead controlled by the size of the particle itself, leading to the estimates  $a^2 {v'}^2$ and $\eps^{2-\gamma}$ for the dimensional and non-dimensional acoustic pressure. Revisiting the arguments of section\,\ref{regimes} about the balance of terms in (\ref{guessdyeq}) with this new estimate for the acoustic pressure gives the following. A distinguished regime involving all the four terms in (\ref{guessdyeq}) is possible and corresponds to $\alpha=1, \, \beta=2$ and $\gamma = -2$. This is the most general regime from which sublimits can be deduced. In particular, for $\gamma=-2\alpha$ and $\beta=2$, the dominant balance is between viscous drag, streaming and acoustic pressure, and our Regime I is recovered.  Similarly, for $\gamma = -2 \alpha$, the balance is between the particle inertia, viscous drag and radiation pressure, analogous to our Regime II. Detailed calculations would however necessary to evaluate the acoustic pressure and assess whether the models we derive for $\gamma \ge 0$ remain unchanged for $\gamma < 0$. 

Other extensions of the present work could include the effects of particle compressibility, heat conduction (both of which have already been accounted for in calculations of acoustic pressure \citep{Dani1986,Doin1996,Doin1997I,Doin1997II,Dani2000,Settnes12}) and slip boundary conditions \citep{Xie2014}. 
The impact of an outer scale $\ell$ of variation of the wave amplitude that differs substantially from the wavelength is relevant to many applications and also deserves consideration. The parameter $k \ell$ measuring this scale discrepancy would need to be included in an extension of the heuristic model (\ref{guessdyeq}) used to assess possible distinguished regimes. Depending on its size relative to $\eps$, new regimes, including a regime involving a four-term balance, will appear. We leave the analysis of these regimes for future work.

\medskip

\noindent
\textbf{Acknowledgements.} Jin-Han Xie acknowledges financial support from the Centre for Numerical Algorithms and Intelligent Software (NAIS).

\appendix

\section{Solution of the Basset equation (\ref{024})} \label{appf}

Applying the Laplace transform to (\ref{024}) gives
\begin{equation}
\begin{aligned}
&M\left(s^2F(s)-sX^{(0)}(0)-\dot X^{(0)} (0)\right) \\
= &-DF(s) -K\left(sF(s)-X^{(0)}(0)\right) - B \left(s^2F(s)-s X^{(0)}(0)-\dot X^{(0)}(0)\right) \sqrt{\pi} s^{-1/2},
\end{aligned}
\end{equation}
where $F(s)=\mathcal{L}\{ X^{(0)} \}$ is the Laplace transform of $X^{(0)}$.
We choose $X^{(0)}(0)=X$ and $\dot X^{(0)}(0)=0$,
so that
\begin{equation}
F(s) = X\frac{Ms+\sqrt{\pi}Bs^{1/2}+K}{Ms^2+\sqrt{\pi}Bs^{3/2}+Ks+D}.
\end{equation}
This function has four poles and a (principal) branch cut associated with $s^{1/2}$. For definiteness, we consider the parameters $M=B=K=D=1$, for which the poles satisfy $s^{1/2}=-1.19496 \pm \ii 0.734487$ and $0.308729 \pm \ii 0.642634$. The second equation leads to two poles, $s_1$ and $s_2$ say, with argument in $(-\pi,\, \pi )$ consistent with the choice of branch cut; the first equation leads to poles on the other Riemann sheet that are irrelevant. 

\begin{figure}
\centering
\includegraphics[width=6cm]{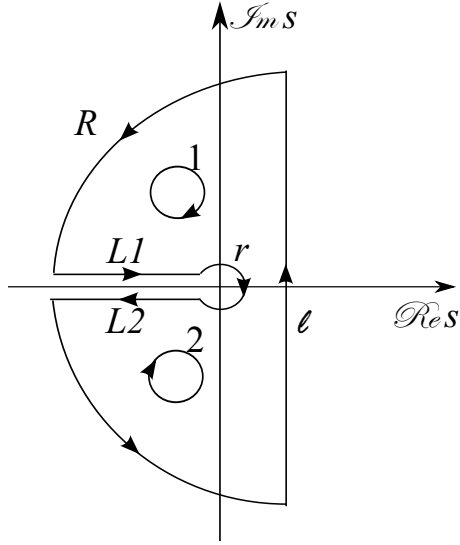}
\caption{Contour for the inverse Laplace transform in (\ref{inlt}).
The countour encloses two poles denoted by 1 and 2, and two sides of the branch cut $(-\infty,0]$ associated with $s^{1/2}$. 
}
\label{fig:contour}
\end{figure}

The particle position $X^{(0)}$ is then obtained by inverting the Laplace transform using the contour  shown in Figure \ref{fig:contour}. This yields
\begin{equation}
\begin{aligned}
\frac{1}{2\pi \ii}\left( \int\limits_{l}^{} + \int\limits_{R}^{} + \int\limits_{r}^{} + \int\limits_{L1}^{} + \int\limits_{L2}^{} \right)  F(s) \e^{st}  \, \d s
 = \textrm{Res}_{1,2}\{ F(s) \e^{st}\}, \label{inlt}
\end{aligned}
\end{equation}
where Res$_{1,2}$ denotes the sum of the residues at $s_1$ and $s_2$.
Taking the limits $R\to\infty$ and $r\to 0$,
$\left( \int\limits_{R}^{} + \int\limits_{r}^{}\right)F(s) \e^{st} \, \d s =0 $, leading to
\begin{equation}
X^{(0)} = \frac{1}{2\pi \ii}\int\limits_{l}^{}F(s) \e^{st} \, \d s =  \textrm{Res}_{1,2}\{ F(s) \e^{st}\} -\frac{1}{2\pi \ii}\left( \int\limits_{L1}^{} + \int\limits_{L2}^{} \right)F(s) \e^{st}\, \d s. \label{a005}
\end{equation}
In (\ref{a005}) $\textrm{Res}_{1,2}\{ F(s) \e^{st}\}$ corresponds to two modes of damped oscillations.  The remaining term gives a continuous spectrum contribution associated with the branch cut; it can be expressed as
\begin{equation}
X^{(0)}_\mathrm{cont} = \frac{-1}{2\pi \ii}\left( \int\limits_{L1}^{} + \int\limits_{L2}^{} \right)F(s) \e^{st} \, \d s =
\frac{-1}{2\pi \ii}\int\limits_{0}^{\infty} \left(f(y)-f(y)^*\right) \, \d y,
\end{equation} 
where 
\begin{equation}
f(y)= \frac{-My+\ii\sqrt{\pi}By^{1/2}+K}{My^2-\ii\sqrt{\pi}By^{3/2}-Ky+D} \e^{-yt}.
\end{equation}


%

\end{document}